# Engineering Waveguide Nonlinear Effective Length via Low Index Thin Films

Wallace Jaffray[1], Farhan Ali[2], Sven Stengel[1], Ziheng Guo[2], Sebastian A. Schulz[2], Andrea Di Falco[2], and Marcello Ferrera[1*]

[1]Institute of Photonics and Quantum Sciences, Heriot-Watt University, SUPA, Edinburgh, EH14 4AS, United Kingdom
[2]School of Physics and Astronomy, University of St. Andrews, Fife, KY16 9SS, United Kingdom
*m.ferrera@hw.ac.uk

**ABSTRACT**

Novel photonic nanowires were fabricated using low-index materials and tested in the near-infrared spectrum to assess their nonlinear optical properties. In this work, we argue the need to redefine the standard nonlinear figure of merit in terms of nonlinear phase shift and optical transmission for a given propagation distance. According to this new metric, our devices largely outperform all established platforms for devices with a linear footprint in the range of 50 to 500 μm, which is demonstrated to be an outstanding technological gap. For 85 fs pulses, with carrier wavelength at 1480nm and sub-μW power levels, a spectral broadening exceeding 80% of the initial bandwidth was recorded over a propagation length of just 50 μm. Leveraging on CMOS-compatible processes and well-established materials such as silicon, silica, and indium tin oxide, our devices bring great promise for developing alternative all-optical devices with unparalleled nonlinear performances within the aforementioned range.

**Introduction**

Integrated nonlinear photonics targets the possibility of controlling light with light over very short propagation distances (of the order of the operational wavelength) and with minimal energy expenditure. The basic motivation behind designing all-optical devices lays on the superior bandwidth of optical modules outperforming the fastest electronics by several orders of magnitude.

The quest for the "perfect" nonlinear material, has recently led to the use of transparent conducting oxides (TCOs) within the near-infrared spectral window, where their real refractive index can become extremely low ($\Re(n)$< 0.2). This condition is associated with a slow light effect leading to giant nonlinearities, which manifest within the ultra-fast optical regime[1,2]. By using near-zero-index (NZI) thin films (thickness < 1 μm) in out-of-plane configurations, ground-breaking nonlinear experiments have been conducted demonstrating record high-frequency conversion in $\chi^{(3)}$ processes[3–5], large bandwidth shifts of the wavelength carrier of an ultra-fast pulse[1,6], and unitary change of the refractive index[7], just to mention a few[8–10]. However, due to the absorptive nature of nonlinearities in TCOs, propagation distances of several microns are already prohibitive, thus rendering impossible the straightforward design of all-optical on-chip devices.

In-plane modulation of optical signals has been achieved via electric bias of transparent electrodes in silicon waveguides[11–14] while optical control has also been explored theoretically[15]. In addition to this, enhanced supercontinuum generation[16] and four-wave mixing[17] have been demonstrated in guided modes evanescently coupled with single-layer materials such as graphene. Other waveguide configurations using 2D materials have also shown optical quenching at ultra-low CW powers when pumped externally[18]. However, these devices operate in the visible range and their temporal response has yet to be reported.

In the present work, a low-index thin film of indium tin oxide (ITO) is deposited on top of silicon nanowires immersed in a FOX cladding on top of a silica substrate. These devices are operated at a central wavelength of 1480 nm corresponding to a real index of 0.97 and imaginary index of 0.63. Nonlinearities are induced on the single propagating mode via evanescent coupling with an upper nonlinear layer of low-index material. In-coupled pulses, with a time duration of 85 fs and power level just below 1μW, showed nonlinear self-broadening exceeding 80% with respect to the input spectrum after a propagation distance of only 50 μm. Propagating losses can be almost entirely ascribed to the coupling into the NZI layer, and they account for about 0.03 dB/μm for a core-to-NZI-layer distance of 170 nm (see Fig. 2 for waveguide design).

Due to the ultra-low power level employed in our tests, we managed to circumvent silicon's nonlinear losses thus allowing for the use of very well-established material platforms. Our study investigates the intrinsic link between a specific device footprint and the most appropriate material for all-optical applications. We also demonstrate that by "diluting" the giant nonlinearities of NZI compounds we can fabricate integrated planar devices with hundreds of microns of propagation distances and unparalleled nonlinear performance for optical modules with equivalent device linear footprint. This offers new alternatives for the design of in-plane all-optical NZI tuneable devices and the realization of optical neural networks and ultra-compact computational photonic units[19,20].

**Nonlinear Material Performance**

To fully appreciate the reported results, we start by considering a set of underpinning materials in integrated photonics and their fundamental nonlinear figure of merit FoM = $n_2/\alpha$, where $n_2$ is the nonlinear Kerr coefficient and $\alpha$ represents the propagation loss coefficient[21]. In Table 1, associated results are reported for the following material systems: graphene[22,23] (representative for 2D materials), gold[24–25] (representative for noble metals), indium tin oxide[6] (ITO) (representative for low-index compounds), silicon nitride[27,28] (representative of nonlinear-absorption-free integrated systems), silicon[29] (most established photonic platform), and silica[30] (cornerstone of long-haul optical data transmission). From the reported data it is immediately apparent how this FoM is limiting and at times misleading in describing nonlinear performances.

For instance, the best nonlinear materials according to Table 1 (highest FoM) are silicon followed by silica glass. The reason for this seemingly superior nonlinear performance refers to the fact that when losses are very low, nonlinear effects can be accumulated over enormous distances thus becoming extremely large in magnitude. However, when the target is an integrated nonlinear chip, kilometres of propagation are not affordable and other material systems must be taken into consideration. Additionally, silicon is affected by other detrimental effects such as two-photon absorption that prevent usability if not at ultra-low power levels. Of course, there are technical solutions to circumvent material limitations and enhance local nonlinearities such as the use of micro-cavities. However, this comes at the price of a largely reduced operational bandwidth while sitting outside the scope of the present manuscript.

To solve the previously mentioned limitations, a more informative and robust figure of merit is needed. This should still take into consideration both losses and nonlinearities but it should also refer to the propagation length associated to the device size. In this regard, we start by considering the infinitesimal nonlinear phase $d\varphi$ accumulated over the infinitesimal propagation length $dz$ defined by $d\varphi = \gamma' I(z)dz$, where: $\gamma' = 2\pi n_2/\lambda$ can be seen as a nonlinear gamma parameter for bulk propagation, and $I(z) = I_0 e^{-\alpha z}$ is the simplified Beer-Lambert law for linear loss propagation. It is worth noting that if we are considering a propagating mode, the effective nonlinear gamma parameter takes into account modal overlap with the nonlinear region. From here we can evaluate the overall nonlinear phase shift for a given propagation length as follows:

$$\phi(L) = \int_0^L \gamma' I_0 e^{-\alpha z} dz = \frac{2\pi n_2 I_0}{\lambda \alpha}(1 - e^{-\alpha L}) \qquad [1]$$

If we now multiply the fraction of remaining intensity after propagation by the accumulated nonlinear phase, we can derive a more descriptive figure of merit:

$$\text{FoM}' = \frac{2\pi n_2 I_0}{\lambda \alpha}(1 - e^{-\alpha L})e^{-\alpha L} \qquad [2]$$

Fig. 1 shows FoM'(L) for the same set of materials in Table 1 and an internal input peak intensity of $I_0$ = 45 GW/cm$^2$ (the one used in our experiment). The coloured bands refer to the lengths at which optical transmissions are between 90% and 10% (See "optical characterization" for extra details on nonlinear parameters evaluation).

| Material | $n_2$ (cm$^2$W$^{-1}$) | $\alpha$ (cm$^{-1}$) | FoM (cm$^3$W$^{-1}$) | $\lambda$ (nm) |
|---|---|---|---|---|
| Si[29] | 4.5×10$^{-14}$ | 3.2×10$^{-8}$ | 1.4×10$^{-6}$ | 1500 |
| SiO$_2$ [30] | 2.7×10$^{-15}$ | 4.6×10$^{-7}$ | 5.8×10$^{-10}$ | 1550 |
| Graphene[22,23] | 1.5×10$^{-9}$ | 9.5×10$^{4}$ | 1.6×10$^{-14}$ | 1550 |
| SiN[27,28] | 2.4×10$^{-15}$ | 2.43 | 9.8×10$^{-16}$ | 1550 |
| ITO[5] | 2.1×10$^{-12}$ | 8.3×10$^{3}$ | 2.6×10$^{-16}$ | 1310 |
| Au[24–26] | 3.1×10$^{-12}$ | 6.7×10$^{5}$ | 5×10$^{-18}$ | 630 |

**Table 1.** Values of Kerr nonlinearity ($n_2$), loss coefficients ($\alpha$), and associated figure of merit (FoM = $n_2/\alpha$) measured at the typical operational wavelength $\lambda$.

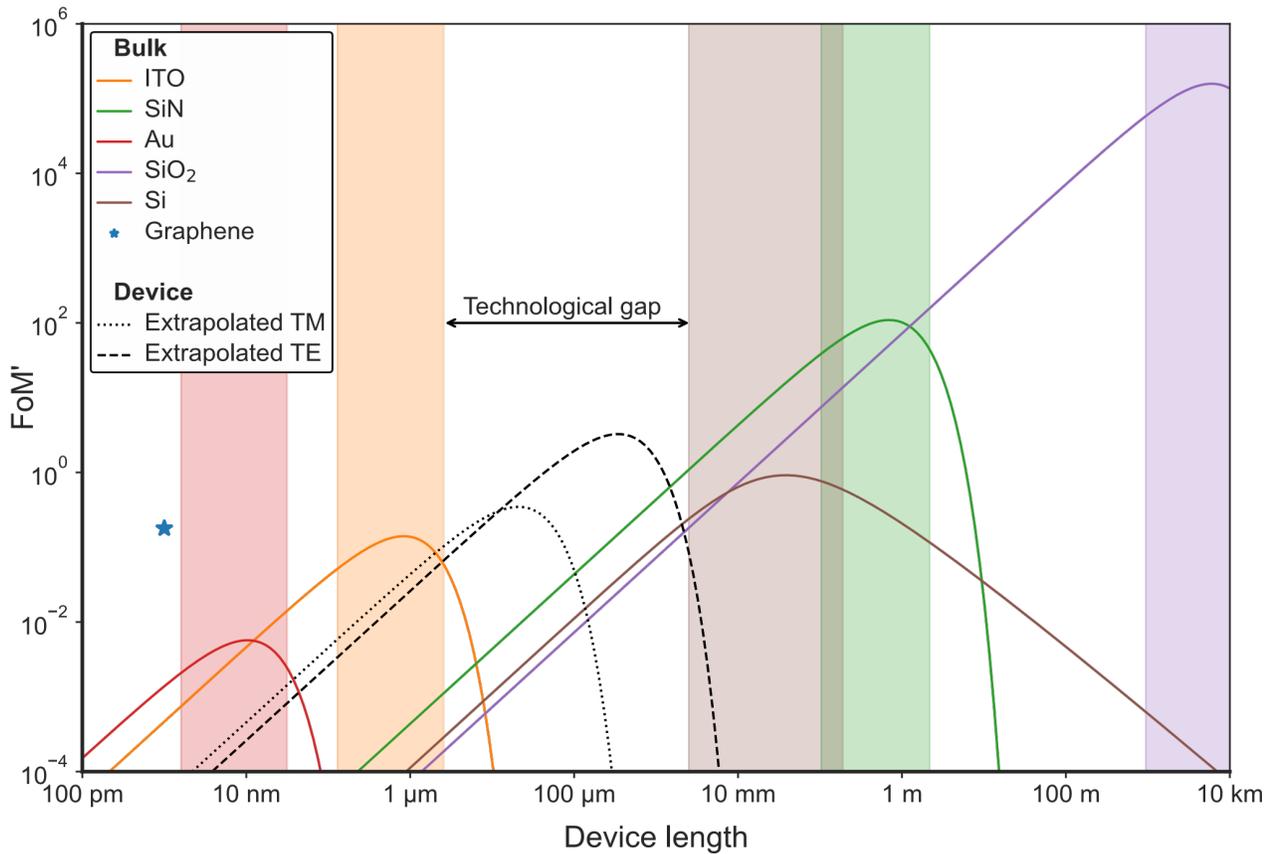

**Figure 1.** Newly defined nonlinear figure of merit (FoM') for different key materials in integrated photonics as a function of a given propagation distance L. FoM' is defined as the total nonlinear phase shift accumulated by a pulse propagating through a material along a distance L and multiplied by the fraction of remaining light after transmission. Table. 1 provides the reference values used to generate this figure. Solid lines and markers represent the bulk materials: Graphene, silicon (Si), silica (SiO$_2$), gold (Au), silicon nitride (SiN), and indium tin oxide (ITO). Dashed lines represent recovered FoM'(L) for the waveguide geometry experimentally tested for both TM and TE modes. All graphs are produced for a power of 45 GW/cm$^2$, which is the power used inside the NZI waveguides in this experiment. The peak value for all curves corresponds to 50% transmission while coloured vertical stripes relates to lengths ranging from 90% to 10% transmission. A direct comparison between the FoM values reported in Tab.1 and the peak values of FoM' in Fig.1 shows the fundamental need to account for optical propagation length to prevent contradictions with common knowledge about optical material properties.

A technological gap for all-optical devices in the range 50 μm -500 μm is apparent, which will be proved to be filled by our nonlinear waveguide technology (*The dashed lines in this figure will be handled later in this work and should be ignored for now*).

From Fig. 1 several interesting considerations can be made. The first and most immediate one is that the picture reflects the typical device length used for a given material. In fact, NZI nonlinearities are best exploited for out-of-plane configurations in thin films with thicknesses below one micron, SiN nonlinear optical devices perform at their best within tens of centimetres of length, and fused silica nonlinear effect are relevant only after several kilometres of propagation (typically considered detrimental for optical communications). In this picture, Graphene is not scalable, and it is represented by one single point for the atomic monolayer. When taking into consideration silicon two-photon absorption by introducing an intensity-dependent $α$[31] the material appears to be outperformed at any given length by another relevant material in strike contrast with the indication provided by the standard FoM.

Finally, arguably the most important feature of this plot is that a material gap is immediately apparent for devices in the range 50 μm -500 μm which can be particularly important for any nonlinear integrated module whose functionalities intrinsically require slightly longer propagation length than what is typically targeted in nanophotonics. An example of this might be the emerging use of optical neural networks for which high nonlinearities are required together with a mode mixing region of the order of 100s μm for telecom wavelength operation[32]. The fundamental question remaining is, how can we fill the previously mentioned technological gap? The following paragraph provides a viable solution to this question.

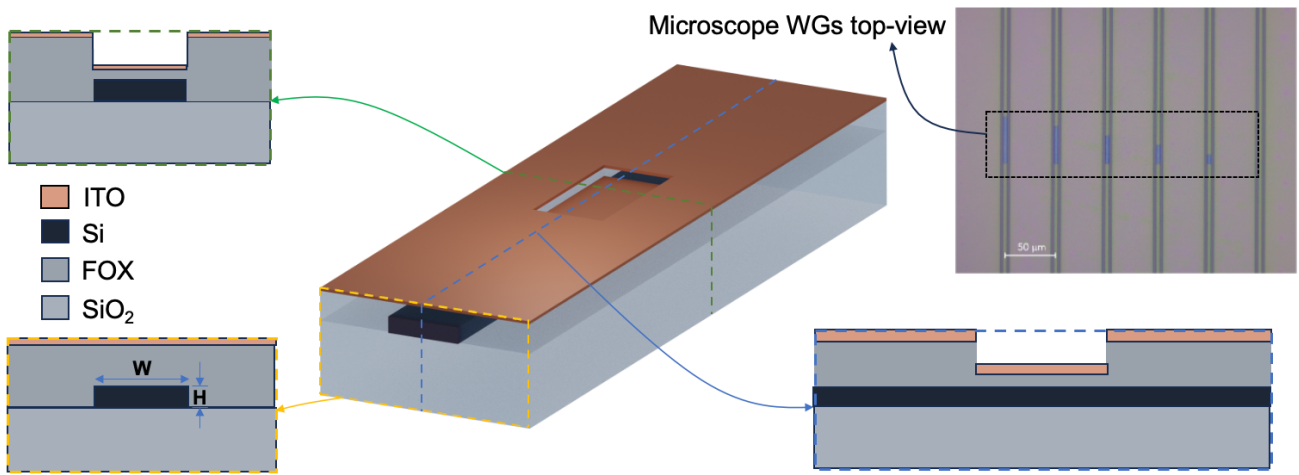

**Figure 2.** Nonlinear waveguide layout (not in scale). Device cross sections are shown at both the input region, where ITO is uncoupled to the propagating mode (bottom left), and at the etched area, where nonlinearities are activated (top left). A side view of the waveguide is given on the bottom right which shows the length of the etched region (nonlinear trench). The top right panel provides an optical microscope image of six ITO-coated nonlinear waveguides with various nonlinear trench lengths including a reference waveguide with no trench at all.

## Nonlinear Waveguide design

One of the key goals of our study is to extend the usability of low-index materials over propagation lengths of a few hundred microns. To do so, our proposed strategy is one of "diluting" optical losses by reducing the field coupling into the lossy nonlinear material. However, since this also comes at the cost of reduced nonlinear effects, we will be fundamentally interested in the effective nonlinear figure of merit of our optical waveguides with lengths of 10s to 100s microns. The geometry of our waveguides is reported in Fig. 2, (not to scale) where a silicon core (W=3um x H=220nm cross section) is surrounded by a FOX cladding to create the fundamental guiding line. On top of this structure lies a 50 nm thick layer of low-index ITO while the substrate is silica. All relevant nonlinear effects take place within and along the etched region, here named as "nonlinear trench" for simplicity. The fabrication process of our devices involved the definition of the waveguides using e-beam lithography, followed by spin coating of flowable oxide (HSQ) on the SOI wafer. For all practical purposes, the spacing layer can be assimilated to $SiO_2$. This was followed by e-beam lithography and dry etching to define the nonlinear trenches.

The final step was the deposition of the upper (nonlinear) ITO layer by e-beam evaporation. As previously mentioned, ultrafast optical nonlinearities are activated by the evanescent tail of the propagating mode, which partially couples into the ITO film. Such coupling can be regulated by varying the size of the silica gap between the waveguide's core and the ITO. To simplify fabrication processes and gain full control over both coupling gap and nonlinear propagation length, etching is performed on the silica cladding before depositing the nonlinear material. Outside the etched region, the ITO film is too far from the core (about 350 nm) to affect mode propagation, as it has been both numerically and experimentally verified. Figure 2 also shows a device cross-section before and at the etched region. The top right panel of Fig. 2 shows an optical microscope image of six waveguides with nonlinear trenches of varying lengths. Our experiments made use of two twin photonic chips with identical layouts, each carrying multiple waveguides with different nonlinear trench lengths, one with the upper ITO layer and one without it. By direct experimental comparison of the two modules, we ensured that at our operational intensities, silicon's nonlinear absorption is negligible, and spurious silicon nonlinearities are avoided.

As the first proof of concept, we decided to choose well-developed material platforms such as silicon and ITO. However, it is important underlying that if we wish to operate above ultra-low power levels to exploit NZI nonlinearities in full, nonlinear absorption-free compounds should be used for the guiding core such as silicon nitride[1]. In addition to this, the low index properties of ITO can also be set almost arbitrarily by either adjusting ITO fabrication parameters[33] or (if an even bigger wavelength shift is required) by using a different TCO[8,10]. Because the optical properties of TCOs can considerably vary with respect to fabrication conditions, for an accurate evaluation of the group velocity dispersion, as well as for the numerical analysis reported in this manuscript, experimentally evaluated dispersion curves for the ITO index were used. The operational wavelength selected for our experiments was at 1480 nm where ITO's real index falls below unity and the losses remain manageable.

Our chip device was tested by using a standard end-fire set-up where light was coupled in and out via two objective lenses (see Fig. 3 for details). A tuneable CW laser (not reported in the picture) was used to characterize the waveguide's linear losses as we coupled light into different waveguides with various nonlinear trench lengths. Our waveguides support both fundamental TE and TM modes, with the TM (E field normal to the multilayer interface) being naturally more "interactive" with the nonlinear ITO layer.

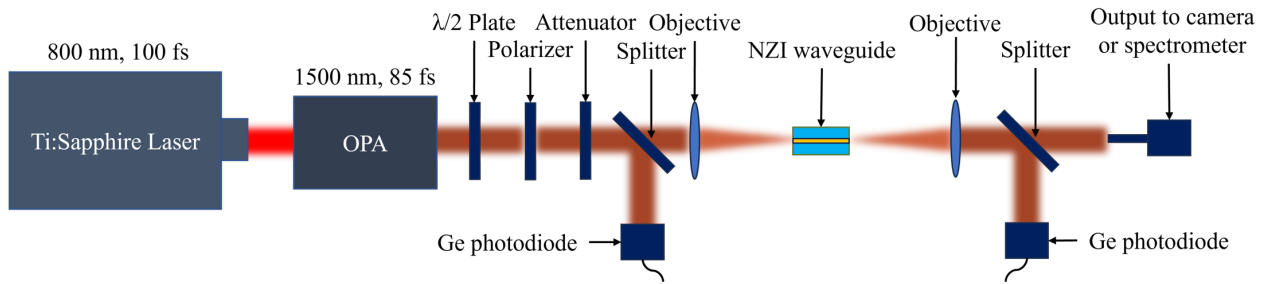

**Figure 3.** Experimental set-up. Devices were characterized by using a standard end-fire apparatus coupled to a femtosecond laser at the input and NIR cameras and photodiodes at the output. A flipping mirror, before the NIR camera (not reported in the figure) allowed the beam to be detoured towards an optical spectrum analyser for the nonlinear spectral analysis.

Propagation losses are experimentally evaluated in Fig. 4 for both TE and TM modes by using input/output power measurements performed on waveguides with different nonlinear trench lengths (3 repetitions of each). The points at length equal to zero are evaluated from unetched waveguides and they are taken as 0 dB loss reference. This allowed for estimation of the differential loss on multiple nonlinear trenches with different lengths (from 10 μm to 50 μm with incremental steps of 10 μm – one additional un-etched waveguide was also included for calibration purposes) and thus allows for the calculation of propagation losses along the nonlinear region with good accuracy. As expected, losses are considerably higher from the TM configuration and match well with numerical simulations. This will be discusses in more details in the "Discussion" session. For the nonlinear characterization, a 1 kHz train of 100 fs pulses centred at 800 nm is used to pump an OPA, generating 85 fs pulses centred at 1480 nm (See Fig. 3 for detailed experimental setup). Input pulses were set to be linearly polarized in either TE or TM polarization, attenuated, and focused by a 60x magnification objective lens onto the facet of the waveguide's core. The output light from our waveguide was collimated via a 40x magnification objective and then sent to a NIR camera, which was used to optimize coupling and alignment. A flipping mirror, before the NIR camera (not reported in the figure) allowed for "detouring" the beam towards an optical spectrum analyser for the nonlinear spectral analysis. Additionally, germanium photodiodes were used in conjunction with calibrated beam samplers to monitor beam stability and coupling conditions before and after the waveguides.

Figure 5 provides an overview of the recorded nonlinear spectral broadening for both TE and TM fundamental modes and is instrumental for describing our strategy when evaluating the waveguides effective Kerr index, which is used to plot dashed device curves in Fig. 1. For an estimated internal input energy of 45 GW/cm$^2$, both TE and TM input spectra were recorded (blue dots figure 5). An interpolation curve of these data was then fed into a Generalized Nonlinear Schrodinger Equation (GNLSE) solver.

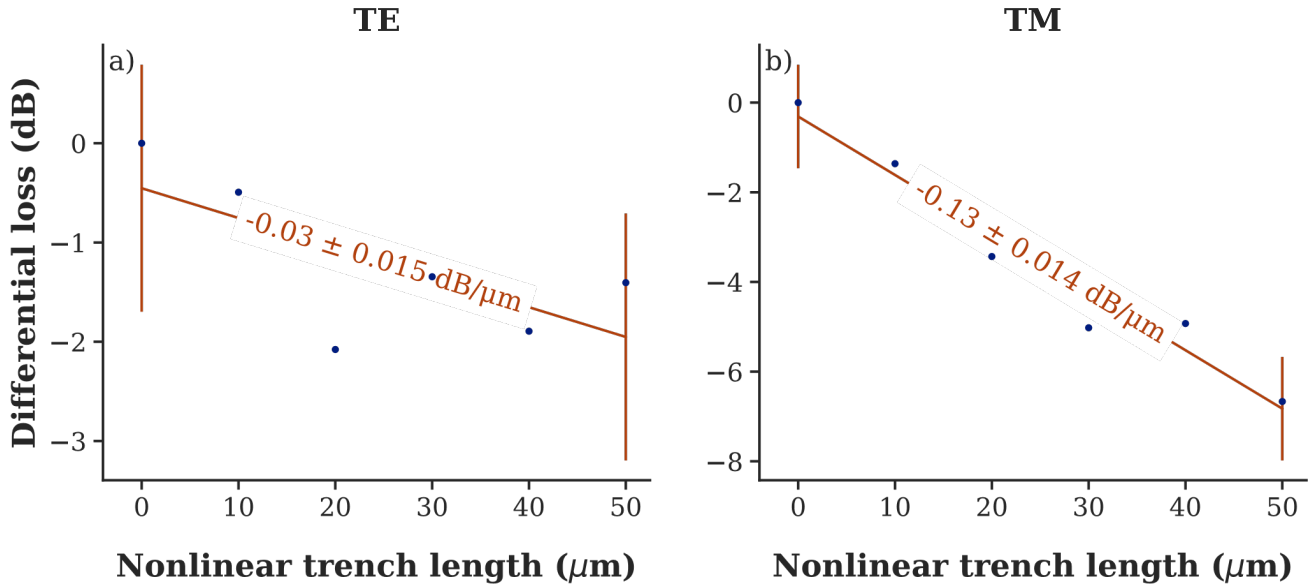

**Figure 4.** Waveguide propagation losses Vs nonlinear trench length for both a) TE and b) TM modes. The points at channel length equal to zero are evaluated from unetched waveguides and they are taken as a 0 dB loss reference.

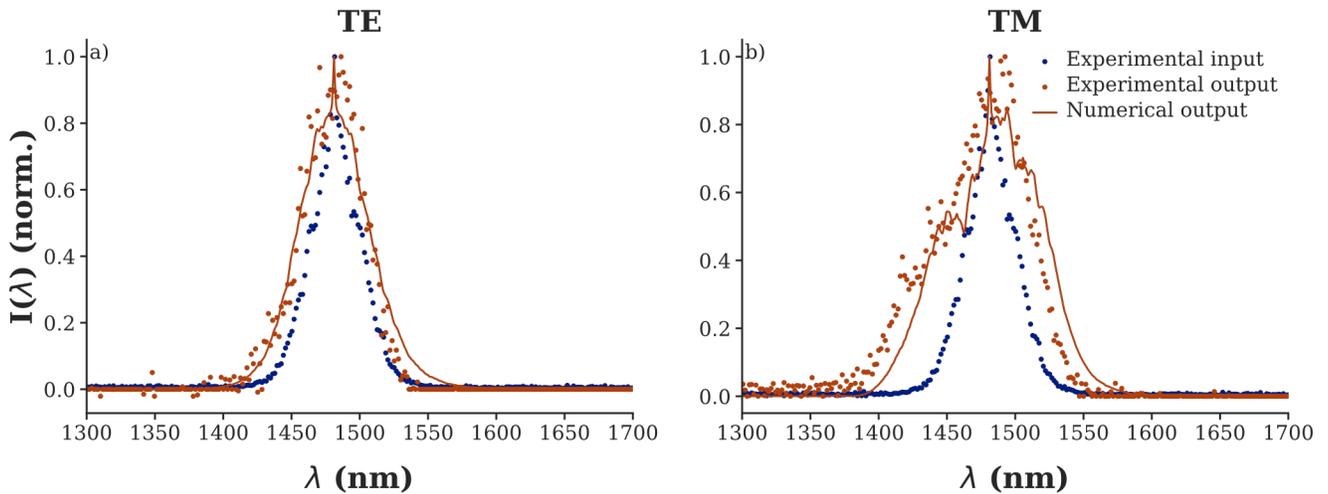

**Figure 5.** Experimentally recorded spectral broadening after $50\mu m$ of propagation length in our NZI-based waveguide measured for both a) TE and b) TM modes. Blue dots provide the pulse input spectrum as recovered from the OPA prior to waveguide coupling. Red dots refer to the waveguides experimental output spectra. Several key parameters are set from initial Eigenmode simulations (loss, dispersion, mode area, and propagation length) then effective nonlinearity parameter $\gamma'$ is estimated by fitting the GNLSE outputs (solid red line) to the experimentally measured output spectra (red dots).

to have a numerical evaluation of the output spectra. The numerical model also made use of additional input parameters such as loss, dispersion, mode area, and propagation length, which were attained via Eigenmode simulations of our waveguides. Finally, the effective gamma parameter of our system is evaluated from fitting our GNLSE's output (solid red lines) to the recorded output spectra. This analysis results in an effective waveguide nonlinearity of $\gamma' = 0.94 \times 10^{-11}$ m/W and $1.6 \times 10^{-11}$ m/W for the TE and TM mode, respectively. The difference in nonlinearity between these modes can be simply understood by looking into the optical power effectively coupled into the ITO layer for both cases. In this regard, a more detailed analysis is reported in the next paragraph.

**Discussion**

Our experimental analysis is carried out on multiple waveguides with the same spacer thickness of 170 nm (in the following referred as nominal spacer thickness - NST) and different lengths for the nonlinear trench region. This waveguide geometry was chosen via a combination of eigenmode simulations supported by preliminary results on material characterisation.

By numerical means, we can also confidently extend our analysis to waveguides with different spacer thicknesses. These studies are reported in Fig. 6, where the top panel (Fig.6-a) shows the losses recovered via eigenmode simulations as a function of spacer thickness. Because our simulations do not account for fabrication imperfections, a lower bound attenuation of 0.01 dB/μm is flatly applied to waveguides for all spacer thicknesses values. This is necessary to prevent the estimated phase shift from diverging (see formula for FoM' at Eq. 1). The value of 0.01 dB/μm is not arbitrary, instead it corresponds to the propagation losses experimentally evaluated from testing our 1 cm long waveguides without the upper ITO layer. As expected, the TE case has much lower losses than the TM case. This logic follows directly from the middle panel (Fig. 6-b), where the power coupled into the ITO layer ($P_C$) is plotted Vs the spacer thicknesses. Here we see how much stronger the TM coupling is with respect to the TE one.

The nonlinear analysis for waveguides with different spacer thicknesses is numerically extrapolated from our experimental results for the NST case. In this regard, we decided to follow a simplified approach according to which the power coupled into the ITO layer is considered proportional to the associated nonlinear gamma parameter. This assumption is reasonable as the power coupled into the ITO layer is relatively low, and thus higher order nonlinearities are not activated. It is also consistent with our experiments as the ratio of the coupled powers of the TE and TM modes (1.83) closely matches the experimentally recovered ratio of TE and TM gamma parameters (1.74) for the tested devices (see green dotted lines in Fig. 6).

In Fig. 6-c we plot the nonlinear phase-shift $\Phi_{MAX}$ evaluated at the maximum of FoM'(L) for different spacer thicknesses for both TE and TM polarisations. The maximum phase shifts are attained via Eq.1 and the correspondent gamma parameters for different spacer thicknesses are retrieved by multiplying normalised active power curves by the correspondent TE and TM gamma parameters for the NST case. The normalisation of the active power curves is performed over the power values for the NST case.

Each point in Fig. 6-c is the peak value of a correspondent FoM'(L) curve like those reported in Fig. 1., which also corresponds to a specific optimal device length for a given spacer thickness. In this regard, since the maximum nonlinear phase shift occurs at a transmission of 50%, the optimal device length as a function of spacer thickness can be directly estimated from the loss curves in Fig. 6-a for a 3dB attenuation. For the sake of direct quantitative evaluation, we report the plot of optimal device length (ODL) Vs the associated spacer thickness in Fig.6-d. It is worth underlying that although the optimal nonlinear propagation length can be set at will by varying the spacer thickness, this would require the fabrication of a new set of devices. However, simply changing the polarization of interest also alters the active power coupled into the nonlinear layer thus changing the FoM'(L) curves whose peak values shift by several tens of microns.

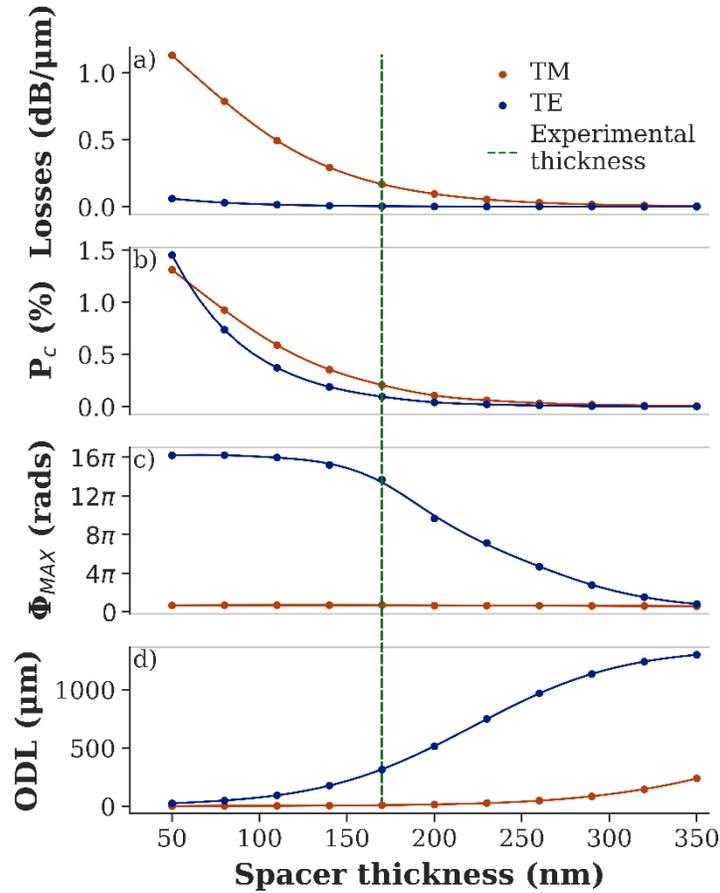

**Figure 6.** Numerical extrapolation of waveguide properties for different optical coupling with the nonlinear layer. Losses, coupled power ($P_C$= percentage of total power in the ITO layer), maximal nonlinear phase ($\Phi_{MAX}$), and optimal device length (ODL) are plotted as a function of spacer thickness for both TE and TM polarisations in panels a), b), c), and d), respectively. The green dashed line highlights the NST points. Plots in panel a) and b) are both attained via Eigenmode simulations. The phase shift in panel c) is found by recording the nonlinear phases evaluated at the maxima of different FOM'(L) curves attained by using Eq.1 for different coupling gaps. Associated gamma parameters for different spacer thicknesses are retrieved by multiplying normalised active power curves by the correspondent TM and TE gamma parameters for the NST case. The normalisation over the active power curves is performed over the power values for the NST case. Each point in panel c) has an associated optimal device length (ODL) which can be attained as guided propagation distance at 3dB attenuation. ODLs values are reported in panel d) for completeness. For all curves in Fig.6 a trend line is plotted together with numerical points to underline the optical behaviour.

## Conclusions

Following the general interest in low index nonlinear technologies and the need for their integration with standard planar photonic modules, we have fabricated and tested a set of nanowires, with record high nonlinear figure of merit (here defined as the product of the nonlinear phase shift acquired along a specific propagation length and the fraction of remaining output light). Our devices exploit the evanescent coupling of the propagating field into an upper nonlinear layer of low index ITO to "dilute" optical losses and provide usable nonlinearities over propagation lengths much longer than the penetration depth in bulk TCOs. This allows for direct control over the nonlinear figure of merit, by simply changing the coupling gap between the guiding core and the nonlinear layer. Ultra-fast pulse propagation (85 fs) in just 50 μm of our optical interconnects lead to a half-π nonlinear phase-shift with a correspondent 80% of spectral broadening. This has been achieved at ultra-low energy levels ($< 1\mu W$), thus allowing to circumvent silicon's intrinsic nonlinear absorption. The proposed device configuration enables arbitrarily setting the optimal nonlinear effective length in the range 50 μm - 500 μm, which is proved to be an outstanding technology gap in current integrated photonics. Filling such a gap might be critical for the further development of specific technologies such as integrated optical neural networks for which the degree of integration cannot overcome the requirements for wiring complexity and computational power.

## Author Contributions

M.F. and A.D.F. conceived the study and supervised the research activity. W.J. and S.S. performed the numerical study and the nonlinear experiments. F.A. fabricated and characterized the linear properties of the devices and of the deposited materials (with support from Z.H. and S.A.S.). All authors contributed to the preparation of the manuscript.

## Funding Sources

The Heriot-Watt team wishes to acknowledge economic support from EPSRC project ID: EP/X035158/1, AFOSR (EOARD) under Award No.FA8655-23-1-7254, and Dstl's Defence Science and Technology Futures (DSTF) programme under DASA contract number ACC6036459. SAS acknowledges support from EPSRC (EP/X018121/1). ADF is supported by the European Research Council (ERC) under the European Union Horizon 2020 Research and Innovation Program (Grant Agreement No. 819346).